\documentclass[aps]{revtex4}
\usepackage{graphicx}
\usepackage{color}
\def\bc{\begin{center}}
\def\ec{\end{center}}

\def\beq{\begin{equation}}
\def\eeq{\end{equation}}

\begin{document}

\title{Quantized dynamics in closed quantum systems}

\author{K. Ziegler}
\affiliation{
Institut f\"ur Physik, Universit\"at Augsburg, D-86135 Augsburg, Germany}
\date{\today}

\begin{abstract}
We propose an approach to process data from interferometric measurements on a closed quantum system
at random times. For this purpose a time correlation matrix is introduced which enables us to extract 
dynamical properties of the quantum system. After defining a generalized expectation value we obtain 
a distribution of time scales, an average transition time and a correlation time. A classical limit
exists which is separated from the quantum fluctuations. The latter are characterized by resonances associated
with poles of the generalized expectation value. Its analytic behavior is studied and some generic 
properties are linked to a quantized Berry phase.  
\end{abstract}

\maketitle


\section{Introduction}

\begin{figure}[t]
\begin{center}
\includegraphics[width=9cm,height=6cm]{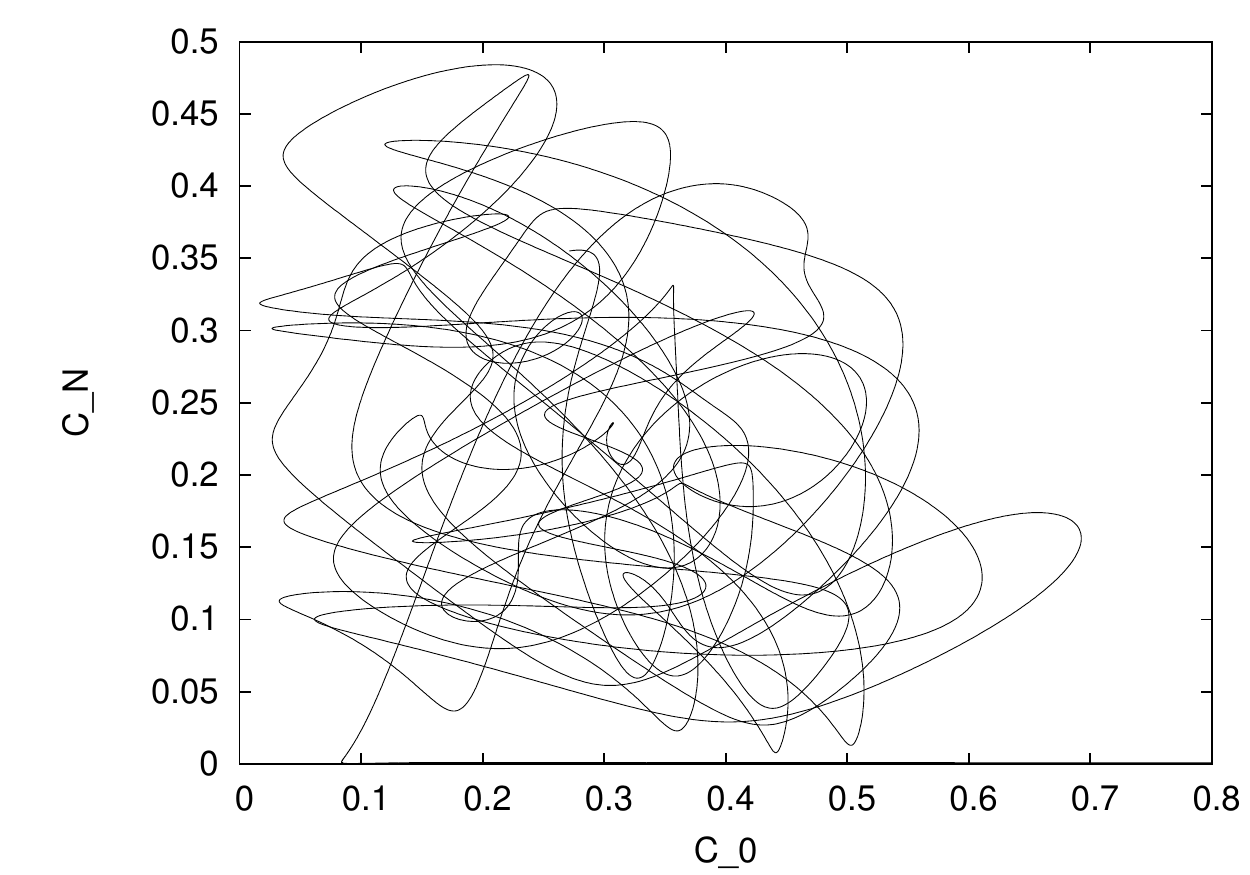}
\caption{
Trajectory of the return $|0,N\rangle\to|0,N\rangle$ vs. transition $|0,N\rangle\to|N,0\rangle$ probability of $N=20$
bosons in a double well (from Ref. \cite{Ziegler_2011}).
}
\label{fig:dynamics}
\end{center}
\end{figure}

Quantum systems have a complex dynamics. Even for a small number of particles this
can be erratic similar to a classical random walk, as indicated in Fig. \ref{fig:dynamics} 
for the tunneling of 20 bosonic particles in a double well potential. Such a behavior suggests that a 
statistical approach would be useful to extract some generic information about the quantum 
evolution. This is supported by the fact that large sets of experimental data are
available whose properties can be treated with statistical concepts.
The idea is not new and found a very successful realization in the Random Matrix Theory (RMT).
It has been applied to many physical systems, such as nuclei, atoms and mesoscopic systems
\cite{10.2307/1970079,PhysRev.104.483,doi:10.1063/1.1703773,1966NucPh..78R.696L,10.2307/2027409,mehta2004random,RevModPhys.69.731}.
The motivation for the RMT is that there is no way of knowing the Hamiltonian of a quantum system
and of calculating the spectrum because of complexity. Instead of a specific Hamiltonian
a random ensemble of Hamiltonians is chosen to describe the physics of a class of systems.
The class is characterized by symmetry of the ensemble with respect to (e.g., orthogonal,
unitary or symplectic) transformations. Here we will pursue a different concept. Except
for the existence of a Hamiltonian with some spectrum and with the overlap of its
energy eigenstates with the initial and the measured state, only the probability of a certain
quantum state at different observation times will enter the statistics. Thus, this dynamical approach is different
from the spectral approach of the RMT, since it is based on the sequence of many independent measurements at
different times on a quantum system with given initial and measured states. 
In more concrete terms, for a given time $t_k$ we evaluate the probability $p_k$ that the
system is in a certain state. Then we compare different discrete times $\{t_1, t_2, ...\}$
and their probabilities $\{p_1, p_2, ...\}$ to calculate statistical properties such as an 
average transition and a correlation time.
Here it is assumed that each experiment is prepared in the same initial state and all
measurements are performed for the same final state. As an example, identical
quantum systems are prepared in a given initial state in a number of independent
laboratories and in each of them a sequence of measurements of a given state 
(i.e., with identical detectors) is repeatedly performed at different times
$t_1<t_2<\cdots <t_k$. For instance, in a double well potential we prepare
a Fock state of $N$ bosons in the right well and let the system evolve. The measurement is
performed by two detectors, one of them clicks when
the left well is empty, the other clicks when the right well is empty. The
probabilities for the measurement as a function of time are visualized by the
trajectory in Fig. \ref{fig:dynamics}. 
The properties of the
distribution of the measurements are determined by the eigenstates and the eigenvalues 
of the Hamiltonian and by the overlap of the initial and the measured states with
all these eigenvectors.

A different application of the RMT has been recently proposed for the description of random measurements.
It is based on Dyson's circular matrix ensemble \cite{doi:10.1063/1.1703773,doi:10.1063/1.1703774,doi:10.1063/1.1703775,mehta2004random}), 
which represents random unitary matrices and has been used as a tool to determine the trace of powers of the density matrix
and the related R\'enyi entropy
\cite{PhysRevLett.108.110503,PhysRevLett.120.050406,PhysRevA.97.023604,PhysRevB.100.134306,PhysRevX.9.031009}.
In contrast to this approach, we consider in the following only random time steps with a specific
Hamiltonian $H$ and given initial and measured states rather than a circular matrix ensemble.

In our approach we perform only one measurement in each individual experiment. This is different from
a monitored evolution, where projective measurements are performed
either periodically~\cite{krovi06,krovi06a,krovi07,Stefanak2008,Gruenbaum2013,luck14,Grunbaum2014,dhar15,sinkovicz16,Friedman_2016,
Friedman2017,thiel18,nitsche18,lahiri19,Yin2019} or randomly \cite{varbanov08,return20,kessler2020detection}.
Although there are some technical similarities between the unitary and the monitored evolution,
the results and their interpretation are quite different in these two cases.

The Berry phase~\cite{berry84} plays an important role in physics as the geometric
phase of the wave function. A classical example is the Aharonov-Bohm effect~\cite{1959PhRv..115..485A},
where the change of the magnetic flux is associated with the change of the Berry phase. Other examples
are the Quantum Hall Effect and topological insulators, where the Berry phase provides, for instance,
a characterization of transport properties. As a general concept, the Berry phase
provides a crucial insight into quantum physics as well as into the physics of classical vector
waves such as electro-magnetic 
waves~\cite{PhysRevLett.100.013904,PhysRevA.78.033834,PhysRevB.80.155103,Ziegler:18}. 
Here we will argue that the Berry phase is also a useful tool for the
characterization of dynamical quantum measurements on finite-dimensional quantum systems.

\section{Time Correlation matrix}
\label{sect:cm}

We consider the transition amplitude from $|\Psi_0\rangle$ to $|\Psi\rangle$
\beq
u_k=\langle\Psi|e^{-iH(\tau_1+\cdots+\tau_k)}|\Psi_0\rangle
\ ,
\eeq
which describes the unitary evolution from the initial state $|\Psi_0\rangle$ and its overlap with 
the state $|\Psi\rangle$ after the time $t_k=\tau_1+\cdots +\tau_k$. In other words, 
$|u_k|^2$ is the probability to find the quantum system in the state $|\Psi\rangle$ after the unitary 
evolution of the initial state $|\Psi_0\rangle$ over the time $t_k$. 
Since the evolution is defined by the Hamiltonian $H$, we consider its eigenstates 
$\{|E_j\rangle\}_{j=1,...,N}$ and its corresponding eigenvalues $\{E_j\}_{j=1,...,N}$ and
write the amplitude in spectral representation as
\beq
u_{k}=\sum_{j=1}^N \langle\Psi|E_j\rangle\langle E_j|\Psi_0\rangle e^{-iE_j(\tau_1+\cdots +\tau_k)}
\equiv \sum_{j=1}^N q_j e^{-iE_j(\tau_1+\cdots +\tau_k)}
\ .
\label{spectr_rep1a}
\eeq
Although the phases are not directly 
accessible experimentally, their effect is observable through the interference of quantum states. This enables us,
for instance, to relate the product of amplitudes at different times with probabilities of interfering amplitudes:
\beq
u_k^*u_{k'}+u_{k'}^*u_{k}
=\frac{1}{2}\left(|u_k+u_{k'}|^2 - |u_k-u_{k'}|^2\right)
\ ,\ \ 
i(u_k^*u_{k'}-u_{k'}^*u_{k})
=\frac{1}{2}\left(|u_k+iu_{k'}|^2 - |u_k-iu_{k'}|^2\right)
\ ,
\eeq
where the probabilities are observable in interferometric measurements. 
%
%
This relation suggests to consider the correlation of the amplitudes at different times $u_k$, $u_{k'}$
through the time correlation matrix (TCM) $\langle u^*_k u_{k'}\rangle_\tau$, where the average $\langle ...\rangle_\tau$
is taken with respect to the distribution of times $\{t_k\}$. For the distribution we assume that $t_k$
consists of $k$ random time steps $\{\tau_k\}$, which are independently and equally distributed.
Then the TCM reads in spectral representation
\[
\langle u_k^*u_{k'}\rangle_\tau=
\sum_{j,j'}q^*_jq_{j'}\langle e^{iE_j(\tau_1+\cdots+\tau_k)}e^{-iE_{j'}(\tau_1+\cdots+\tau_{k'})}\rangle_\tau
\]
\beq
=\sum_{j,j'}q^*_jq_{j'}\cases{
\langle e^{i(E_j-E_{j'})(\tau_1+\cdots+\tau_k)}\rangle_\tau\langle e^{-iE_{j'}(\tau_{k+1}+\cdots+\tau_{k'})}\rangle_\tau & for $k'>k$ \cr
\langle e^{i(E_j-E_{j'})(\tau_1+\cdots+\tau_{k'})}\rangle_\tau\langle e^{iE_{j'}(\tau_{k'+1}+\cdots+\tau_{k})}\rangle_\tau & for $k'<k$ \cr
\langle e^{i(E_j-E_{j'})(\tau_1+\cdots+\tau_{k})}\rangle_\tau & for $k'=k$ \cr
}
\ .
\label{product_ua}
\eeq
Defining $\lambda_j=\langle e^{iE_j\tau}\rangle_\tau$ and $\lambda_{jj'}=\langle e^{i(E_j-E_{j'})\tau}\rangle_\tau$
this becomes
\beq
\langle u_k^*u_{k'}\rangle_\tau=\sum_{j,j'}q^*_jq_{j'}\cases{
\lambda_{jj'}^k\lambda_{j'}^{k'-k}& for $k'\ge k$ \cr
\lambda_{jj'}^{k'}{\lambda^*_{j}}^{k-k'}& for $k'<k$ \cr
}
=\sum_{j,j'}q^*_jq_{j'}\cases{
(\lambda_{jj'}/{\lambda}_{j'})^k{\lambda}_{j'}^{k'}& for $k'\ge k$ \cr
(\lambda_{jj'}/{\lambda^*_{j}})^{k'}{\lambda^*_{j}}^{k}& for $k'<k$ \cr
}
\ .
\label{u_product}
\eeq
The TCM decays exponentially with $|k-k'|$, provided $|\lambda_j|<1$. 
For fixed $|k-k'|$ the TCM is constant for the diagonal elements $\lambda_{jj}=1$, 
though. This reflects the fact that a unitary evolution between the same energy eigenstates
gives just a phase factor $e^{-iE_j\tau}$ (cf. Eq. (\ref{spectr_rep1a})). On the other hand, 
these phase factors lead to a decay of different energy states due to interference effects after
the time average. The diagonal element $\langle |u_k|^2\rangle_\tau$ is the probability to
measure the state $|\Psi\rangle$ at time $t_k$. It has the asymptotic behavior
\beq
\langle |u_k|^2\rangle_\tau
=\sum_{j,j'=1}^Nq^*_jq_{j'}\lambda_{jj'}^k
\sim P_N:=\sum_{j=1}^N|q_j|^2
\label{diag_elements}
\eeq
for $k\sim\infty$. The asymptotic probability $P_N$ 
stores important information regarding the
properties of the quantum system. Since $|q_j|^2=|\langle\Psi|E_j\rangle|^2|\langle\Psi_0| E_j\rangle|^2$
is the product of the overlaps between the energy eigenstate $|E_j\rangle$ with the initial state
and with the measured state, it provides a measure of how much this energy eigenstate contributes to the
transition $|\Psi_0\rangle\to |\Psi\rangle$ during the unitary evolution. For instance, the asymptotic behavior
of the return probability to the initial state $|\Psi_0\rangle\to |\Psi_0\rangle$ with the dimensionality $N$ of
the underlying Hilbert space characterizes Anderson localization when $\lim_{N\to\infty}P_N>0$ and the
absence of Anderson localization if $\lim_{N\to\infty}P_N= 0$ \cite{cohen16}. This can be understood when
we note that the normalization implies $\sum_{j=1}^N|\langle\Psi_0| E_j\rangle|^2=1$ and that for a localized state
only a few energy eigenstates have a nonzero overlap with $|\Psi_0\rangle$. For a delocalized
state, on the other hand, the overlap is nonzero for a large number of energy eigenstates which is
of the order of $N$. An extreme case is given when these overlaps are equal. Then we have 
$|\langle\Psi_0| E_j\rangle|^2=1/N$ due to the normalization, which implies $P_N=1/N$.
Another example would be diffusion, where $P_N$ also vanishes for $N\sim\infty$.

Besides the diagonal elements we can also extract relevant information from the off-diagonal
elements of the TCM, which depend on the phases of the  $\lambda$'s. 
The off-diagonal TCM elements decay quickly with $|k-k'|$ according
to Eq. (\ref{u_product}), nevertheless, we can identify long- and short time scales.
Then the scale of the decay provides a measure of how strongly the quantum dynamics is correlated in time.
With $\lambda_j=|\lambda_j|e^{i\varphi_j}$ and with the asymptotic behavior $\lambda_{jj'}\sim\delta_{jj'}$ 
we get from Eq. (\ref{u_product}) 
\beq
\langle u_k^*u_{k'}\rangle_\tau\sim C_{k-k'}:=\sum_{j}|q_j|^2|\lambda_j|^{|k-k'|}e^{i\varphi_j(k'-k)}
\ .
\label{off-diagonals}
\eeq
In the remainder of this paper we will analyze the statistical properties of the TCM.

\section{Summary of the main results}
\label{sect:summary}

Starting from $C_{k-k'}$ of Eq. (\ref{off-diagonals}) we consider the average
transition time
\[
\langle k\rangle_z=\frac{\sum_{k,k'\ge 1}|z|^{k+k'}k'C_{k-k'}e^{i\omega(k'-k)}}
{\sum_{k,k'\ge 1}|z|^{k+k'}C_{k-k'}e^{i\omega(k'-k)}}
\]
with $z=|z|e^{i\omega}$. With this expression we also get the average correlation time
as $\langle k-k'\rangle_z= 2i Im(\langle k\rangle_z)$. The average transition time separates
into a classical part $1/(1-|z|^2)$ and a quantum part. The latter is characterized by
$3N$ poles in the $N$--dimensional Hilbert space. These poles can be seen as resonances 
of the quantum dynamics. Then we use a Laurent expansion of $\langle k\rangle_z$ in powers of $z$
and identify its coefficients with the characteristics of the quantum dynamics. It turns out that
$\langle k\rangle_z$ is related to the Berry connection and the coefficient ${\tilde c}_0$
of the Laurent expansion is related to the corresponding Berry phase. This enables us to
extract some universal properties that do not depend on the details of the underlying quantum 
model.
 
\section{Fluctuating quantum dynamics}
\label{sect:statistics}

The elements of the Hermitean TCM in Eq. (\ref{off-diagonals}) can be considered as complex ``weights'' of 
a statistical ensemble of discrete times $\{k-k'\}$.
For large $|k-k'|$ the decaying behavior of the TCM is determined by the largest value(s) of $|\lambda_j|$.
The phase factor also plays an important role in the statistics of the unitary evolution.
To analyze the effect of the phases we consider the Fourier transformation of $C_{k-k'}$ 
in Eq. (\ref{off-diagonals}) (cf. Eq. (\ref{FT00}) in App. \ref{app:summation})
\beq
C(z^*,z')
:=\sum_{k,k'\ge 1}C_{k-k'}{z^*}^k{z'}^{k'}
=\frac{z^*z'}{z^*z'-1}\sum_{j=1}^N|q_j|^2\frac{|\lambda_j|^2 z^*z'-1}{(z'\lambda_j-1)(z^*\lambda^*_j-1)}
\ ,
\label{gen_function00}
\eeq
which becomes for $z'=z$
\beq
C(z^*,z)
=\frac{|z|^2}{|z|^2-1}\sum_{j=1}^N|q_j|^2\frac{|\lambda_j|^2|z|^2-1}{|z\lambda_j-1|^2}
\ .
\label{gen_function01}
\eeq
Next we consider the differential expression
\beq
K(z^*,z'):=z'\partial_{z'}\log[C(z^*,z')]
=\frac{\sum_{k,k'\ge 1}{z^*}^k{z'}^{k'}k'\sum_{j}|q_j|^2|\lambda_j|^{|k-k'|}e^{i\varphi_j(k'-k)}}
{\sum_{k,k'\ge 1}{z^*}^k{z'}^{k'}\sum_{j}|q_j|^2|\lambda_j|^{|k-k'|}e^{i\varphi_j(k'-k)}}
\ ,
\label{appr_case00}
\eeq
which reads with the parametrization $z=|z|e^{i\omega}$ and $z'=|z|e^{i\omega'}$
\[
=\frac{\sum_{k,k'\ge 1}|z|^{k+k'}k'\sum_{j}|q_j|^2|\lambda_j|^{|k-k'|}e^{i(\varphi_j+\omega)(k'-k)}e^{i(\omega'-\omega)k'}}
{\sum_{k,k'\ge 1}|z|^{k+k'}\sum_{j}|q_j|^2|\lambda_j|^{|k-k'|}e^{i(\varphi_j+\omega)(k'-k)}e^{i(\omega'-\omega)k'}}
\ .
\]
Here $0<|z|<1$ guarantees that the summation with respect to $k'=k$ exists.
$|z|$ can be understood as a parameter that suppresses TCM elements at large times.
This can be corrected at the end of the calculation by sending $|z|\to1$.
As we will see below, the limit $|z|\to1$ exists for $\log(|z|)K(z^*,z')$.
On the other hand, $|z|<1$ is not unrealistic in the sampling of the TCM
elements in numerical or experimental situations, since very large numbers of measurements
are less likely than smaller numbers to be sampled. From this point of view, $|z|<1$ mimics the
limitations of accessible large time TCM elements and can be understood as a smooth projection
onto a shorter times scale. 

By setting $z'=z$ the expression of $K(z^*,z')$ leads to the connection
\beq
{\cal C}(z):=K(z^*,z)
=\frac{\sum_{k,k'\ge 1}k'\sum_{j}|q_j|^2|\lambda_j|^{|k-k'|}e^{i\varphi_j(k'-k)}z^{k'-k}|z|^{2k}}
{\sum_{k,k'\ge 1}\sum_{j}|q_j|^2|\lambda_j|^{|k-k'|}e^{i\varphi_j(k'-k)}z^{k'-k}|z|^{2k}}
\ ,
\label{connection00}
\eeq
where we have used $z^*=|z|^2/z$. Returning to $z=|z|e^{i\omega}$ we get
\[
{\cal C}(z)=\frac{\sum_{k,k'\ge 1}|z|^{k+k'}k'\sum_{j}|q_j|^2|\lambda_j|^{|k-k'|}e^{i(\varphi_j+\omega)(k'-k)}}
{\sum_{k,k'\ge 1}|z|^{k+k'}\sum_{j}|q_j|^2|\lambda_j|^{|k-k'|}e^{i(\varphi_j+\omega)(k'-k)}}
\ ,
\]
which defines an average transition time $\langle k\rangle_z$ with respect to
\[
P_{k'}=\frac{\sum_{k\ge 1}|z|^{k+k'}\sum_{j}|q_j|^2|\lambda_j|^{|k-k'|}e^{i(\varphi_j+\omega)(k'-k)}}
{\sum_{k,k'\ge 1}|z|^{k+k'}\sum_{j}|q_j|^2|\lambda_j|^{|k-k'|}e^{i(\varphi_j+\omega)(k'-k)}}
\]
with a real denominator as normalization. For the TCM we define a correlation time 
$-i\langle k-k'\rangle_z=2Im{\cal C}(z)$.
The name connection is chosen because ${\cal C}(z)$ will turn out to be related to a Berry phase
through an integral over a closed contour.
The effect of the Fourier transformation is a global shift of the phases $\varphi_j\to\varphi_j+\omega$.
This implies that a changing $\omega$ can measure the phase sensitivity of $\langle k\rangle_z$. 
We can also create higher moments of $k-k'$ from the $\omega$--derivatives of $2iIm [{\cal C}(z)]$.
From this point of view, the connection ${\cal C}(z)$ is a generating function for moments of $k-k'$.
It is obvious that the connection ${\cal C}(z)$ diverges in the limit $|z|\to 1$,
since the system returns under unitary evolution to any of the accessible states in the Hilbert
space after some finite time interval, regardless of the total time. However, we will see that for 
long times this repetitive behavior separates as a classical process from fluctuating quantum behavior.

\subsection{Evaluation of the connection}
\label{sect:conn}

The sum on the right-hand side of Eq. (\ref{gen_function00}) can be rewritten
with a common denominator as
\beq
\sum_{j=1}^N|q_j|^2\frac{|\lambda_j|^2 z^*z'-1}{(z'\lambda_j-1)(z^*\lambda^*_j-1)}
=\frac{1}{Q(z^*)Q'(z')}\sum_{n,n'=0}^{N} g_{n,n'}{z^*}^nz'^{n'}
\label{exp2}
\eeq
with
\[
Q(z^*)=\prod_{j=1}^N (z^*\lambda^*_j-1)
\ ,\ \ \
Q'(z')=\prod_{j=1}^N (\lambda_j z'-1)
\ .
\]
Rewriting the sum with $z'=z+\Delta$ and $z^*=|z|^2/z$ we get a polynomial in $z$ of order $2N$
besides an extra factor $z^{-N}$. This allows us to write the polynomial as a product:
\beq
z^{-N}\sum_{n=0}^N\sum_{n'=0}^{N}|z|^{2n}g_{n,n'}z^{N-n}(z+\Delta)^{n'}=z^{-N}g_{0,N}(z-z_1)\cdots(z-z_{2N})
\ ,
\label{gen_function02}
\eeq
where $\{z_m\}$ are the zeros that depend on $\Delta$ and on $|z|$. In other words, when we fix $|z|$ as an extra
parameter we obtain complex numbers $\{z_m\}$ that satisfy Eq. (\ref{gen_function02}) for any value of $z$.
Thus, the connection is a function defined on the complex plane:
\beq
{\cal C}(z)
=z\partial_{z'}\log[C(z^*,z')]\Big|_{z'=z}
=\frac{1}{1-|z|^2}-z\left(\sum_{j=1}^N\frac{1}{z-1/\lambda_j}
+\sum_{n=1}^{2N}\frac{z_n'}{z-\bar{z}_n}\right)
\ ,
\label{connection}
\eeq
where $\{\bar{z}_n\}$ are the zeros of the expression $C(z^*,z)$ 
in Eq. (\ref{gen_function01}) at fixed $|z|$. 
Moreover, according to Eq. (\ref{coeff0}) we have
\beq
z_n'=\partial_\Delta z_n\Big|_{\Delta=0}
=-\frac{\sum_{m=0}^N\sum_{m'=0}^{N}|z|^{2m}g_{m,m'}m'(\bar{z}_n)^{N-m+m'-1}}
{g_{0,N}\prod_{m\ne n}(\bar{z}_n-\bar {z}_m)}
\ .
\label{diff_z}
\eeq
$z_n'$ can diverge when $\bar{z}_n$ is getting closer to another pole 
$\bar {z}_m$. This is the case, for instance, when the $\{\lambda_j\}$ are very close to each
other.

\section{Discussion}

Now with ${\cal C}(z)$ in Eq. (\ref{connection}) we are in a position to return to the connection in 
Eq. (\ref{connection00}) and to study its analytic properties. 
For instance, we can set $z=|z|$ to obtain
\[
{\cal C}(|z|)
=\frac{\sum_{k,k'\ge 1}k'C_{k-k'}|z|^{k'+k}}{\sum_{k,k'\ge 1}C_{k-k'}|z|^{k'+k}}
=\frac{1}{1-|z|^2}-\sum_{j=1}^N\frac{1}{|z|-1/\lambda_j}
-\sum_{n=1}^{2N}\frac{z_n'}{|z|-\bar{z}_n}
\ .
\]
The poles of ${\cal C}(z)$ play the role of resonances of the quantum system, where the connection diverges and the
average transition time is very long. Even when $z$ does not reach these
poles, the average transition time and the average correlation time $2Im [{\cal C}(z)]$ increase in their vicinity.
Through this effect the resonances characterize the quantum dynamics. 
The resonances also have an amplitude, which is either one or $z_m'$, according to Eq. (\ref{connection}).
For a closer inspection of the resonances the connection ${\cal C}(z)$ in Eq. (\ref{connection}) can 
be expanded in a Laurent series as
\beq
{\cal C}(z)=\sum_{n\ge0}{\tilde c}_n(z-z_0)^n+\sum_{n>0}{\tilde c}_{-n}(z-z_0)^{-n}
\ ,
\label{laurent}
\eeq
where the coefficients are expressed as Cauchy integrals
with respect to a contour $\gamma$ in a region where ${\cal C}(z)$ is analytic:
\[
{\tilde c}_n=\frac{1}{2\pi i}\int_{\gamma}{\cal C}(z)(z-z_0)^{-n-1}dz
\ \ \ (-\infty<n<\infty) 
\ .
\]
Some information regarding the resonances is already provided by the coefficient Laurent ${\tilde c}_{0}$:
When we choose the contour $\gamma$ such that it encloses $z_0$ and all $2N$ poles $\{\bar{z}_m\}$ but only
$N'$ from the $N$ poles $\{1/\lambda_j\}$ we get
\beq
{\tilde c}_0=\frac{1}{2\pi i}\int_\gamma{\cal C}(z)\frac{1}{z-z_0}dz
=\frac{1}{1-|z|^2}+N-N'-\sum_{j}\frac{z_0}{z_0-1/\lambda_j}
\ ,
\label{c_0}
\eeq
where the last sum includes only the $N-N'$ poles which are not enclosed by $\gamma$.
It is remarkable that ${\tilde c}_0$ depends only on some $\lambda$'s.
All other coefficients depend on the model parameters $\{\lambda_j\}$ and $\{\bar{z}_m\}$.
With $z=|z|e^{i\omega}$ and $z_0=0$ the Laurent series in Eq. (\ref{laurent}) is a Fourier series of the connection
${\cal C}(z)$, where the coefficients ${\tilde c}_n|z|^n$ represent the spectral 
weights of the frequency $n\omega$.

\subsection{Properties under scaling}

The scaling transformation $\lambda_j\to t\lambda_j$ implies for the connection in Eq. (\ref{connection00})
\beq
{\cal C}_t(z)
:=\frac{\sum_{k,k'\ge 1}k'\sum_{j}|q_j|^2|t\lambda_j|^{|k-k'|}e^{i\varphi_j(k'-k)}z^{k'-k}|z|^{2k}}
{\sum_{k,k'\ge 1}\sum_{j}|q_j|^2|t\lambda_j|^{|k-k'|}e^{i\varphi_j(k'-k)}z^{k'-k}|z|^{2k}}
\ .
\label{sc_conn}
\eeq
Moreover, for small $t$ we get for the Fourier transform of the TCM in Eq. (\ref{gen_function00})
\[
C_t(z^*,z')=\frac{z^*z'}{z^*z'-1}\sum_{j}|q_j|^2[1+t(\lambda_j^*z^*+\lambda_j z')] + O(t^2)
\]
such that the connection also reads
\beq
{\cal C}_t(z)=z\partial_{z'}\log C_t(z^*,z')\Big|_{z'=z}
\sim \frac{1}{1-|z|^2}+\frac{\sum_j|q_j|^2 t\lambda_jz}{\sum_j|q_j|^2[z+ t(\lambda_j z^2+|z|^2\lambda_j^*)]}
\ ,
\label{sc_conn1}
\eeq
where the second term vanishes for $t\to0$. We can also take the limit $t\to 0$ directly for the TCM 
$t^{|k-k'|}C_{k-k'}\to \delta_{k,k'}$ and calculate the connection from Eq. (\ref{sc_conn}) as
\[
{\cal C}_{t=0}(z)
=\frac{\sum_{k\ge 1}k|z|^{2k}}{\sum_{k\ge 1}|z|^{2k}}=\frac{1}{1-|z|^2}
\ ,
\]
which can be considered as a classical limit of our model. It describes a 
system that evolves from an initial position to a final position. The probability to reach
the final position after discrete time steps $k=1,2,...$ decreases with increasing time
according to $p_k=|z|^{2k}/\sum_{k\ge 1}|z|^{2k}$, since the system moves away from its initial 
position. It should be noted that this classical limit is also obtained 
when we fix $|z|$ and send only $z\to\infty$:
\[
\lim_{z\to\infty}{\cal C}(z)=\frac{1}{1-|z|^2}-N-\sum_{n=1}^{2N}z_n'
=\frac{1}{1-|z|^2}
\]
due to $\sum_nz_n'=-N$ in Eq. (\ref{cond11}).

Next we consider small quantum fluctuations around the classical limit.
For this purpose we write for Eq. (\ref{sc_conn1})
\[
{\cal C}_t(z)\sim
\frac{1}{1-|z|^2}+\frac{z}{(z-z_+)(z-z_-)}
\ , \ \ z_{+/-}=-\frac{{\bar\mu}}{2t\mu}\pm\sqrt{\frac{{\bar\mu}^2}{4t^2\mu^2}-|z|^2\frac{\mu^*}{\mu}}
\ ,
\]
where $\mu=\sum_j|q_j|^2\lambda_j$ and ${\bar\mu}=\sum_j|q_j|^2$.
When the contour $\gamma$ encloses both poles we get the classical result. On the other hand,
when $\gamma$ encloses only $z_+$ but not $z_-$, we get the Cauchy integral
\beq
{\tilde c}_0=\frac{1}{2\pi i}\int_\gamma{\cal C}_t(z)\frac{1}{z}dz
\sim\frac{1}{1-|z|^2}+\frac{1}{z_+ -z_-}
\sim \frac{1}{1-|z|^2}-t|z|^2\frac{\mu^*}{{\bar\mu}}
\ .
\label{small_t}
\eeq
This result indicates that the average transition time and the average correlation time are
changed due to quantum fluctuations as
\beq
\frac{1}{2\pi}\int_0^{2\pi}\langle k\rangle_z d\omega
\sim \frac{1}{1-|z|^2}
-t|z|^2\frac{\sum_j|q_j|^2\lambda_j^*}{\sum_j|q_j|^2}
\ ,\ \  
\frac{i}{2\pi}\int_0^{2\pi}\langle k-k'\rangle_z d\omega
\sim 2t|z|^2\frac{\sum_j|q_j|^2 Im(\lambda_j)}{\sum_j|q_j|^2}
\ .
\eeq
When $\gamma$ encloses only $z_-$ but not $z_+$ the contribution of the quantum fluctuations
would appear with the opposite sign.


\subsection{Connection and Berry phase}

The integer $N$ in the Laurent coefficient ${\tilde c}_0$ of Eq. (\ref{c_0}), which is the dimensionality
of the underlying Hilbert space, originates in a topological invariant through a quantized Berry phase.
The reason is that the connection ${\cal C}(z)$ is related to the Berry connection \cite{berry84}
\[
{\cal A}(z)=-i\partial_{\omega'}\frac{\langle{\hat u}^*(z){\hat u}(z')\rangle}
{\sqrt{\langle |{\hat u(z)}|^2\rangle}\sqrt{\langle|{\hat u}(z')|^2\rangle}}
\Big|_{z'=z}
\ ,
\]
where $\langle ...\rangle$ refers here to the usual quantum expectation value. 
Then the Berry phase $w_\gamma$ reads
\[
w_\gamma
=\frac{1}{2\pi}\int_0^{2\pi}{\cal A}(z)d\omega
=\frac{1}{2\pi i}\int_\gamma{\cal A}(z)\frac{1}{z}dz
\ ,
\]
and since $\sqrt{\langle|{\hat u}(z')|^2\rangle}$ in the denominator of the Berry connection
is a real single valued function whose differential does not contribute to the loop integral, 
the Berry phase becomes
\beq
w_\gamma=\frac{1}{2\pi i}\int_\gamma \frac{-i\langle{\hat u}^*(z)\partial_{\omega}{\hat u}(z)\rangle}
{\langle |{\hat u(z)}|^2\rangle}\frac{1}{z}dz
=\frac{1}{2\pi i}\int_\gamma-i\partial_{\omega}\log[\langle{\hat u}^*(z){\hat u}(z')\rangle]\Big|_{z'=z}
\frac{1}{z}dz
\ .
\eeq
Therefore, the integral in Eq. (\ref{c_0}) is the corresponding Berry phase of the connection ${\cal C}(z)$. 
This can also be understood when we consider the expression (\ref{connection}) after substracting the classical
term $1/(1-|z|^2)$. Enclosing all poles by $\gamma$ we find
\beq
\frac{1}{2\pi i}\int_\gamma
\left[\sum_{j=1}^N\frac{1}{z-1/\lambda_j}
+\sum_{n=1}^{2N}\frac{z_n'}{z-\bar{z}_n}\right]dz=0
\ ,
\eeq
which holds regardless of the parameters of the model. What looks like a trivial result originates
in a topologically invariant Berry phase $w_\gamma=N$ of the connection
\[
{\cal C}_1(z)
=-i\partial_{\omega'}\log\left[\sum_{n,n'=0}^{N} |z|^{2n}g_{n,n'}z^n {z'}^{n'}\right]\Big|_{\omega'=0}
=-i\partial_{\omega'}\log\left[(z-z_1)\cdots(z-z_{2N})\right]\Big|_{\omega'=0}
\]
\[
=-z\sum_{n=1}^{2N}\frac{z_n'}{z-\bar{z}_n}
\ ,
\]
provided that all poles are enclosed by the contour $\gamma$.
Since $\sum_nz_n'=-N$ due to Eq. (\ref{cond11}) we get
\beq
\frac{1}{2\pi i}\int_{\gamma}{\cal C}_1(z)\frac{1}{z}dz=N
\ .
\eeq
Thus, the Berry phase of ${\cal C}_1(z)$ 
is an invariant that depends on the dimensionality of the underlying Hilbert space but is
independent of the other model details.

\subsection{Example: symmetric two-level system}
\label{sect:example}

\begin{figure}[t]
\begin{center}
\includegraphics[width=7cm,height=5cm]{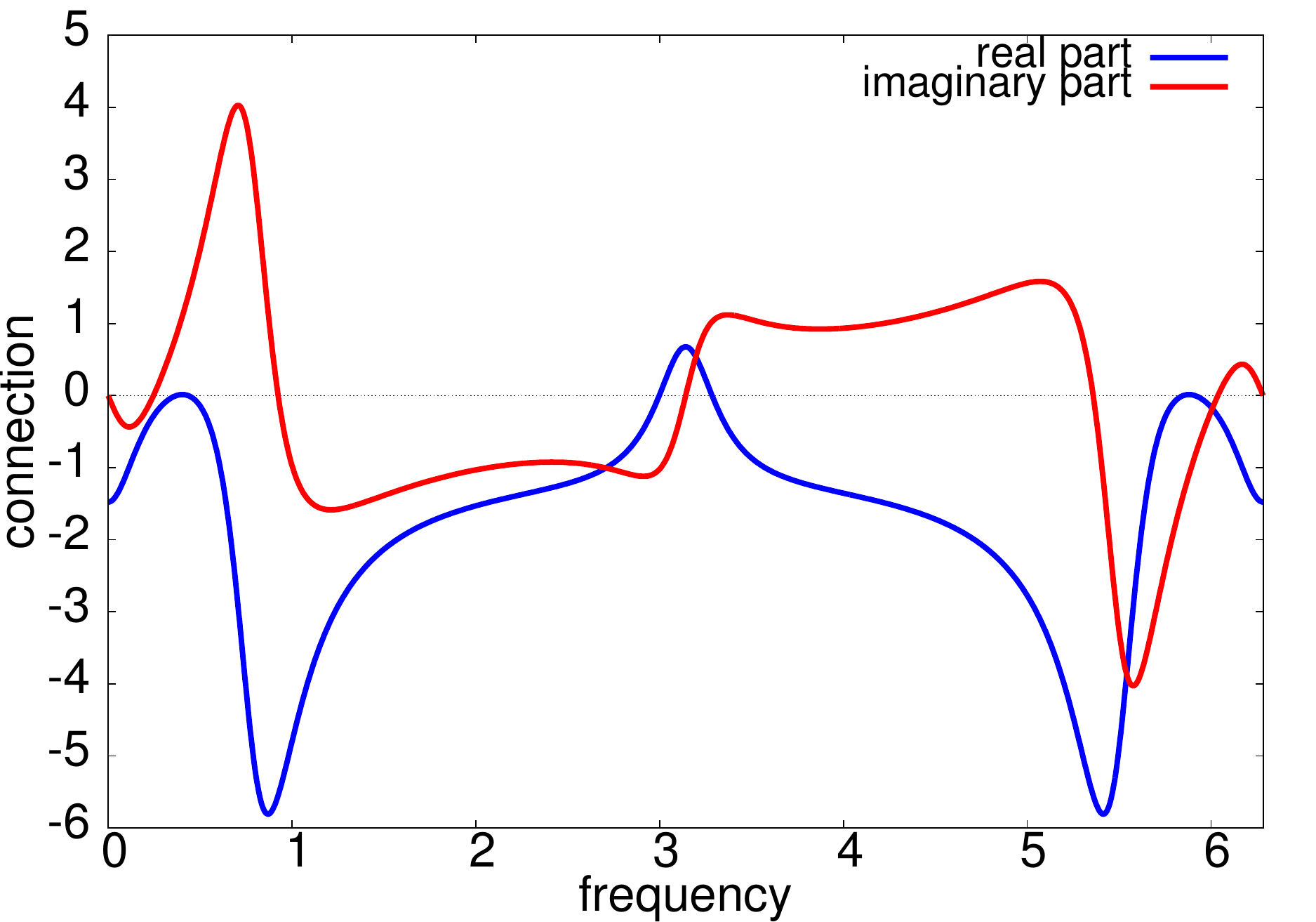}
\caption{
The connection ${\cal C}(e^{i\omega})-1/(1-|z|^2)$ of a symmetric two-level system with $\lambda_1=(1+i)/\sqrt{3}$ 
for fixed overlaps $|q_1|=|q_2|$. There are four resonances at $1/\lambda_1$, $1/\lambda_1^*$ and $\pm1/|\lambda_1|$.
}
\label{fig:connection}
\end{center}
\end{figure}

As a special application of the TCM approach we consider $N=2$, i.e., a two level
system. Moreover, we assume a symmetric case for simplicity with energy levels $\pm J$,
which implies $\lambda_2=\lambda_1^*$, and with overlaps $|q_1|=|q_2|$.
Then the $3\times3$ coefficient matrix of Eq. (\ref{exp2}) reads
\[
(g_{n,n'})=|q_1|^2\pmatrix{
0 & -|\lambda_1|^2(\lambda_1+\lambda_1^*) & 2|\lambda_1|^4 \cr
\lambda_1+\lambda_1^* & 0 & -|\lambda_1|^2(\lambda_1+\lambda_1^*) \cr
-2 & \lambda_1+\lambda_1^* & 0\cr
}
\ .
\]
The connection is simple for $|z|=1$ with all poles outside the unit circle:
\[
\bar{z}_1=1/\lambda_1^*
\ ,\ \  
\bar{z}_2=-|\lambda_1|^{-1}
\ ,\ \ 
\bar{z}_3=|\lambda_1|^{-1}
\ ,\ \ 
\bar{z}_4=1/\lambda_1
\ .
\]
From Eq. (\ref{diff_z}) we get get identical coefficients $z_1'=\cdots =z_4'=-1/2$,
which implies $z_1'+z_2'+z_3'+z_4'=-2$ in agreement with the general relation of Eq. (\ref{cond11}).
The connection reads
\[
{\cal C}(z)
=\frac{1}{1-|z|^2}-\frac{1}{2}\left(
\frac{z}{z-1/\lambda_1}+\frac{z}{z-1/\lambda_1^*} -\frac{z}{z-1/|\lambda_1|}-\frac{z}{z+1/|\lambda_1|}
\right)
\]
and is visualized without the classical term $1/(1-|z|^2)$ in Fig. \ref{fig:connection} 
on the unit circle for $\lambda_1=(1+i)/\sqrt{3}$.
The Laurent coefficient ${\tilde c}_0$ reads for the symmetric two-level system
\beq
{\tilde c}_0=\frac{1}{2\pi i}\int_\gamma{\cal C}(z)\frac{1}{z}dz
=\frac{1}{1-|z|^2} +K_\gamma
\ ,
\eeq
where the quantum contribution $K_\gamma=0, \pm 1/2,\pm 1$ depends on the choice of $\gamma$.

\section{Conclusions}
\label{sect:conclusions}

Sequences of interferometric measurements on a closed quantum systems at different times can be collected in terms
of the TCM. This Hermitean matrix has been used to define a generalized expectation value with
respect to a distribution of transition and correlation times. The average transition
time is a sum of simple poles, where the latter represent characteristic resonances of the
quantum dynamics. This enabled us to extract some universal features through the associated
Berry phase. More detailed properties of the measured quantum system can be systematically derived
from the coefficients of the Laurent series of the average transition time.

\section*{Acknowledgments:} 
I would like to thank Eli Barkai who introduced me to the physics of repeated measurements in quantum systems
and David Kessler for enlightening discussions .
The support of this work by the Julian Schwinger Foundation is gratefully acknowledged.

\appendix

\section{Summations over discrete time}
\label{app:summation}

We consider summations of the type given by the Fourier transformation in Eq. (\ref{gen_function00}):
\[
\sum_{k,k'\ge 1;k\ge k'}a^k b^{k'}
=\sum_{k\ge 1}a^k\sum_{k'=1}^k b^{k'}
=\sum_{k\ge 1}a^kb\frac{1-b^{k}}{1-b}
=\frac{b}{1-b}\left[\sum_{k\ge 1}a^k-\sum_{k\ge 1}(ab)^k\right]
\]
\[
=\frac{b}{1-b}\left[\frac{a}{1-a}-\frac{ab}{1-ab}\right]
=b\frac{a(1-ab)-ab(1-a)}{(1-a)(1-b)(1-ab)}
=b\frac{a-ab}{(1-a)(1-b)(1-ab)}
=\frac{ab}{(1-a)(1-ab)}
\ .
\]
With this we obtain
\beq
\sum_{k,k'\ge 1;k\ge k'}\left(a^k b^{k'}+c^k d^{k'}\right)-\frac{ab}{1-ab}
=\frac{ab(ac-1)}{(a-1)(ab-1)(c-1)}
\label{rel5}
\eeq
when $d=ab/c$.

Now we consider the specific case with
\[
a= e^{-i\omega}\lambda^*_j
\ ,\ \ 
b=e^{i\omega'}/\lambda^*_j
\ ,\ \ 
c=e^{i\omega'}\lambda_j
\ ,\ \ 
d=e^{-i\omega}/\lambda_j
\ ,
\]
which apparently obeys the condition $d=ab/c$, and consider
\[
c_j(e^{-i\omega},e^{i\omega'})
=\sum_{k,k'\ge1;k\ge k'}(\lambda^*_je^{-i\omega})^k(e^{i\omega'}/\lambda^*_j)^{k'}
+\sum_{k,k'\ge1;k'\ge k}(e^{-i\omega}/\lambda_j)^k(e^{i\omega'}\lambda_j)^{k'}
-\sum_{k\ge 1}e^{i(\omega'-\omega)k}
\]
\[
=\sum_{k,k'\ge1;k\ge k'}\left[(\lambda^*_je^{-i\omega})^k(e^{i\omega'}/\lambda^*_j)^{k'}
+(e^{i\omega'}\lambda_j)^{k}(e^{-i\omega}/\lambda_j)^{k'}\right]
-\sum_{k\ge 1}e^{i(\omega'-\omega)k}
\]
which becomes with relation (\ref{rel5})
\[
\frac{|\lambda_j|^2e^{2i\omega'}e^{-2i\omega}-e^{i\omega'}e^{-i\omega}}
{(e^{i\omega'}\lambda_j-1)(\lambda^*_je^{-i\omega}-1)(e^{i(\omega'-\omega)}-1)}
\ .
\]
Moreover, with $z=e^{i\omega}$, $z'=e^{i\omega'}$ we get
\beq
c_j(z^*,z')=\frac{z^*z'(|\lambda_j|^2z^*z'-1)}
{(z'\lambda_j-1)(\lambda^*_jz^*-1)(z'z^*-1)}
\ ,
\label{FT00}
\eeq
which yields $C(z^*,z')=\sum_{j=1}^N|q_j|^2c_j(z^*,z')$ in Eq. (\ref{gen_function00}).
In particular, for $\omega'=\omega$ we obtain
\[
c_j(z^*,z)=\frac{|\lambda_j|^2-1}
{(e^{i\omega}\lambda_j-1)(\lambda^*_je^{-i\omega}-1)(e^{i(\omega'-\omega)}-1)}
\ .
\]


\section{Relation of the zeros}
\label{sect:relation}

Starting from Eq. (\ref{gen_function02})
we take the derivative with respect to $\Delta$ 
and send $\Delta\to0$. This gives 
\beq
\sum_{n=0}^N\sum_{n'=0}^{N}|z|^{2n}g_{n,n'}n'z^{N-n+n'-1}=-g_{0,N}\sum_{m=1}^{2N}z_m'\prod_{n\ne m}(z-\bar {z}_n)
\ ,
\label{diff1}
\eeq
where we have used $z_n=\bar{z}_n+\Delta z_n'+O(\Delta^2)$. This is an equation for a 
polynomial of order $2N-1$ in $z$. Comparing the coefficients of $z^{2N-1}$ yields 
\[
g_{0,N}N=-g_{0,N}\sum_{m=1}^{2N-1}z_m'
\ ,
\]
which implies the relation
\beq
\sum_{m=1}^{2N-1}z_m'=-N
\ .
\label{cond11}
\eeq
Setting $z=\bar{z}_l$ in Eq. (\ref{diff1}) gives
\[
\sum_{n=0}^N\sum_{n'=0}^{N}|z|^{2n}g_{n,n'}n'(\bar{z}_l)^{N-n+n'-1}
=-g_{0,N}z_l'\prod_{n\ne l}(\bar{z}_l-\bar {z}_n)
\ ,
\]
which implies
\beq
z_l'=-\frac{\sum_{n=0}^N\sum_{n'=0}^{N}|z|^{2n}g_{n,n'}n'(\bar{z}_l)^{N-n+n'-1}}
{g_{0,N}\prod_{n\ne l}(\bar{z}_l-\bar {z}_n)}
\ ,
\label{coeff0}
\eeq
provided that there is no degeneracy of $\{\bar{z}_n\}$.

Another useful relation is based on
\[
\partial_{z'}\log\left(\sum_{n,n'=0}^N|z|^{2n}g_{n,n'}z^{N-n} {z'}^{n'}\right)\Big|_{z'=z}
=\frac{\sum_{n,n'}|z|^{2n}g_{n,n'}n'z^{N-n+n'-1}}{\sum_{n,n'}|z|^{2n}g_{n,n'}z^{N-n+n'}}
=-\sum_{m=1}^{2N}\frac{z'_m}{z-\bar{z}_m}
\ ,
\]
where the second equation is due to Eq. (\ref{gen_function02}).
Now using a Laurent expansion around $z=0$ we can compare the corresponding coefficients of powers of $z$,
which gives for each power $z^l$ a relation between the coefficients $\{g_{n,n'}\}$ on one side and
$\{\bar{z}_n\}$, $\{z_n'\}$ on the other.

\bibliography{ref}
\bibliographystyle{iopart-num}

\end{document}